\documentclass[conference]{IEEEtran}

\ifCLASSINFOpdf
\else
\fi
\PassOptionsToPackage{hyphens}{url} 

\usepackage{tikz}
\usepackage{amsmath}

\usepackage{xcolor}
\usepackage{jslisting}
\usepackage{subfigure}
\usepackage{float}
\usepackage{url}
\usepackage{amssymb} 
\usepackage{pifont} 

\usepackage{hyperref}
\hypersetup{
  colorlinks   = true,    
  urlcolor     = blue,    
  linkcolor    = blue,    
  citecolor    = red      
}
\usepackage{multirow}

\newcommand\vvv{Manifest V3}

\newcommand{\onlineExtensions}{200,000}

\newcommand\maliciousExtensions{517}

\newcommand\extensionsOurDataset{289,000}

\newcommand\stillRunFirst{154} 
\newcommand\stillRunFirstPercentage{29.8\%} 
\newcommand\stillRunSecond{290}
\newcommand\stillRunSecondPercentage{56\%}

\newcommand\notStillRunFirstPercentage{70.2\%} 

\newcommand\war{\textit{web\_accessible\_resources}}

\begin{document}
%
\title{Work-in-Progress: Manifest V3 Unveiled: Navigating the New Era of Browser Extensions}

\author{
  \IEEEauthorblockN{Nikolaos Pantelaios}
  \IEEEauthorblockA{North Carolina State University \\ npantel@ncsu.edu}
  \and
  \IEEEauthorblockN{Alexandros Kapravelos \\}
  \IEEEauthorblockA{North Carolina State University \\ akaprav@ncsu.edu}
}

\IEEEoverridecommandlockouts
\makeatletter\def\@IEEEpubidpullup{6.5\baselineskip}\makeatother
\IEEEpubid{\parbox{\columnwidth}{
    {\fontsize{7.5}{7.5}\selectfont Workshop on Measurements, Attacks, and Defenses for the Web (MADWeb) 2024 \\
        1 March 2024, San Diego, CA, USA \\
        ISBN 979-8-9894372-2-1 \\
        https://dx.doi.org/10.14722/madweb.2024.23080 \\
        www.ndss-symposium.org}
  }
  \hspace{\columnsep}\makebox[\columnwidth]{}}

\maketitle

\begin{abstract}

  Introduced over a decade ago, Chrome extensions now exceed \onlineExtensions\ in number. In 2020, Google announced a shift in extension development with Manifest Version 3 (V3), aiming to replace the previous Version 2 (V2) by January 2023. This deadline was later extended to January 2025. The company's decision is grounded in enhancing three main pillars: privacy, security, and performance. This paper presents a comprehensive analysis of the Manifest V3 ecosystem. We start by investigating the adoption rate of V3, detailing the percentage of adoption from its announcement up until 2024. Our findings indicate, prior to the 2023 pause, less than 5\% of all extensions had transitioned to V3, despite the looming deadline for the complete removal of V2, while currently nine out of ten new extensions are being uploaded in Manifest V3. Furthermore, we compare the security and privacy enhancements between V2 and V3 and we evaluate the improved security attributable to V3's safer APIs, examining how certain APIs, which were vulnerable or facilitated malicious behavior, have been deprecated or removed in V3. We dynamically execute 517 confirmed malicious extensions and we see a 87.8\% removal of APIs related to malicious behavior due to the improvements of V3. We discover that only \stillRunFirst\ (\stillRunFirstPercentage) of these extensions remain functional post-conversion. This analysis leads to the conclusion that V3 reduces the avenues for abuse of such APIs. However, despite the reduction in APIs associated with malicious activities, the new Manifest V3 protocol is not immune to such behavior. Our research demonstrates, through a proof of concept, the adaptability of malicious activities to V3. After the proof of concept changes are applied, we showcase \stillRunSecond\ (\stillRunSecondPercentage) of the examined malicious extensions retain their capability to conduct harmful activities within the V3 framework. They can achieve this by incorporating web accessible resources, a method that facilitates the injection of third-party JavaScript code. Conclusively, this paper also pioneers by documenting the impact of user and community feedback in the transition from V2 to V3, analyzing the percentage of initial issues that have been resolved, and proposing future directions and mitigation strategies for the continued evolution of the browser extension ecosystem.

\end{abstract}

\section{Introduction} \label{sec:introduction}

Browser extensions have become an integral component of today’s web browsing experience, offering a range of functionalities that enhance user productivity and security. The typical browser user often has multiple extensions installed, spanning various categories such as ad blockers like uBlock Origin~\cite{unblockOrigin}, password managers such as LastPass~\cite{passwordManager}, and productivity tools like BlockSite~\cite{blockSite}. These extensions are of particular interest to the web security community due to their access to high-privilege APIs. These APIs enable extensions to modify web pages and bypass the browser’s Same Origin Policy (SOP), among other capabilities. Consequently, a persistent challenge for webstore platforms is distinguishing between benign and malicious extensions and effectively mitigating the latter before they impact users.

\begin{figure*}[ht]
  \centering
  \includegraphics[width=7in,height=3in,keepaspectratio]{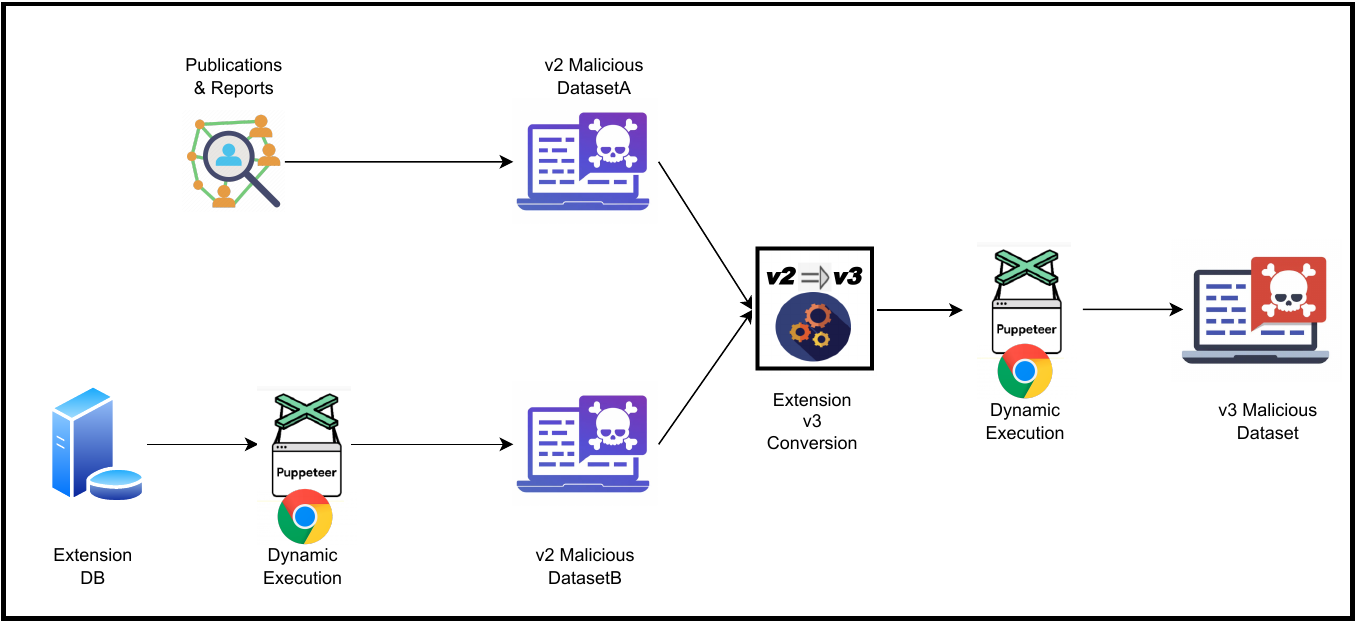}
  \caption{A full description of our architecture including the data collection process, the conversion to Manifest V3 and the dynamic analysis stage.}
  \label{fig:newArchitecture}
\end{figure*}

To tackle the existing problem of extensions violating user's privacy and browser security, Google announced in 2020 changes in extensions design, reflected by upgrading Manifest version 2 (Manifest V2 or just V2) to Manifest version 3 (Manifest V3 or just V3). Google's transition from Manifest V2 to V3, initially planned for early 2023 but delayed to address developer concerns, aimed to enhance Chrome's extension ecosystem's security, privacy, and performance. V3 faced criticism for potentially impacting ad blockers and other extensions, leading to a more nuanced rollout. In contrast, Firefox implemented V3 differently, focusing on preserving ad-blocker functionality. As of late 2023 and early 2024, Google continued refining V3 based on community feedback, aiming to balance security improvements with the functionality desired by extension developers and users.

The transition to V3 has sparked diverse reactions, particularly among developers. Key changes in V3 include the discontinuation of third-party code inclusions and the replacement of certain APIs, which, while aimed at enhancing security, have raised concerns about potential compatibility issues and limitations in extension functionality. One of the most notable changes is the shift from the \textit{webRequest} API to \textit{declarativeNetRequest}, which, despite aiming for improved performance and privacy, has introduced complexities and bugs, potentially impacting the security of V3. Additionally, the new protocol limits extension collaboration, posing challenges for privacy-related extensions which now face restrictions in filter categories and request rates. These changes, while designed to streamline extension capabilities, have inadvertently limited their functionality and interoperability. Furthermore, the need for privacy-related extensions to maintain dynamic blocklists within the extension code has introduced additional overhead and frequent updates, possibly affecting browser performance and the web developer experience. V3's categorization encompasses remote code inclusion, API changes, content security policy (CSP) rules, and the introduction of service workers. Notably, the prohibition of third-party JS code inclusion and the alterations in API usage, like the restricted usage of \textit{xhr} and \textit{executeScript}, signify a considerable shift in extension development. The stringent CSP rules in V3, particularly the deprecation of the \emph{unsafe-eval} rule, further dictate the security landscape, restricting arbitrary code execution. Lastly, replacing background scripts with service workers aims at enhancing browser performance, though it introduces complexities in extension behavior and performance dynamics.

In our research, we extensively analyze the Manifest V3 ecosystem, focusing on the adoption rate and the security and privacy enhancements over V2. Our study involves converting V2 malicious extensions to V3 and dynamically analyzing them. Notably, we find that V3's implementation results in the removal of 87.8\% of APIs related to vulnerable or malicious behavior, demonstrating a significant increase in security. However, our dynamic testing reveals that some extensions still manage to exhibit malicious behavior in V3, highlighting the need for ongoing vigilance in the extension ecosystem despite the considerable security improvements.

Our study compiles a total of \maliciousExtensions\ malicious V2 extensions from dynamic analysis and prior research. We then adapt these extensions to V3, following rule-based conversions outlined in official documentation. Our dynamic testing reveals that \stillRunFirst\ (\stillRunFirstPercentage) of these extensions remain functional post-conversion. Moreover, \stillRunSecond\ (\stillRunSecondPercentage) extensions can function in V3 with a \war\ declaration in the manifest file. The full extent of our architecture can be found in Figure~\ref{fig:newArchitecture} and we open source our dataset in Github \footnote{\href{https://github.com/anonymousSubmitterSudo/malicious_v2_v3_extensions}{Link to Github}}.

Our paper makes the following contributions:
\begin{itemize}
  \item We showcase the improvement of V3 against V2, with 87.8\% of the APIs related to malicious or vulnerable JavaScript code not being available in V3 due to deprecation or replacement of those APIs
  \item We present a proof of concept to include 3rd-party resources in extensions developed in Manifest V3 to study malicious code that can still exist in the new ecosystem and propose mitigations
  \item We present the first study on the evolution and the adoption rate of the new Manifest V3 extension design over the past four years it was announced
  \item We open-source our historical dataset with the malicious extensions in both the Manifest V2 and Manifest V3 versions for future comparisons and to kickstart future research on this domain
\end{itemize}

\section{Background} \label{sec:bg}

\subsection{Manifest V3 Community Collaboration} \label{sec:bg:quotes}

Google's development of Manifest V3 focuses on enhancing the security, privacy, and performance of browser extensions, with promises of improvements acknowledged in their statement and supported by sources like Ghostery~\cite{ghostery_study, manifestv3_1}. Google's iterative improvements to V3, such as the transition to the \textit{declarativeNetRequest} API and adjustments to cosmetic rules filters, demonstrate their commitment to refining the extension platform while addressing developer concerns. Efforts to maintain and enhance extension capabilities, alongside a responsive strategy to community feedback, show Google's dedication to evolving the extension ecosystem to meet security and privacy standards effectively.

\subsection{Categorizing Manifest V3 Changes} \label{sec:bg:v3}

\noindent\textbf{Code Inclusion} A key change in \vvv\ is the prohibition of third-party JavaScript (JS) code inclusion, targeting extensions that inject third-party code. Only code bundled with the extension is permitted, with \war\ declarations allowing for limited third-party interactions.

\noindent\textbf{API Changes} \vvv\ introduces API modifications like replacing \textit{webRequest} with \textit{declarativeNetRequest} and restricting \textit{xhr} usage, enhancing security and performance. The \textit{executeScript} API moves to \textit{scripting}, and several APIs are deprecated or replaced to improve security~\cite{declarativeNetRequest, content_script, apiDocumentation}.

\noindent\textbf{CSP Rules} \vvv\ tightens CSP rules, notably deprecating \emph{unsafe-eval} and narrowing permissible CSP values to enhance security.

\noindent\textbf{Service Workers} The shift to \textit{service workers} from \textit{background scripts} aims to boost performance and align with browser architectures, emphasizing asynchronous, event-driven operations~\cite{service_workers}.

\subsection{Extensions Structure} \label{sec:bg:structure}

The source code of an extension is comprised of a combination of JavaScript (JS), HTML, CSS files, media files, and JSON files. The central configuration file, known as \textit{manifest.json}, specifies the extension's permissions. This manifest file delineates all the JS files associated with the extension. In Manifest V2, these JS files are categorized as content scripts and background scripts. However, in Manifest V3, the background scripts have been replaced by service workers. The key distinction between these lies in the set of APIs accessible to each script and the context in which the script operates.

Following the specification of all the JS and media files used by the extension, it is submitted to the webstore for approval. Once accepted, the extension is assigned a unique 32-byte hash identifier, referred to hereafter as the 'id'.

\subsection{Tools} \label{sec:bg:tools}

\begin{table}
    \centering
    {\small

        \begin{tabular}{c|c} \hline
            \multirow{2}{*}{\textbf{Malicious category}} & \textbf{Have source code} \\
                                                         & \textbf{(multi-labels)}   \\ \hline
            Click scam                                   & 33                        \\ \hline
            Ad replacement                               & 112                       \\ \hline
            User data analytics                          & 356                       \\ \hline
            Credentials stealing                         & 3                         \\ \hline
            Browser modification                         & 111                       \\ \hline
            Other                                        & \multirow{2}{*}{90}       \\
            (crypto, mining, gambling, phishing)         &                           \\ \hline
            Total (unique)                               & 517                       \\ \hline
        \end{tabular}
    }
    \caption{Malicious extensions categories based on manual labeling. Each malicious extension can have multiple labels.}
    \label{tab:malCategories1}
\end{table}

\textit{Playwright}~\cite{playwright} is a browser automation tool that operates in a secure and isolated environment. \textit{Playwright} is capable of simulating a variety of browsers, including Chrome, Firefox, Opera, Safari, and Edge. While our analysis currently focuses solely on Chrome extensions, we utilize Playwright to facilitate potential future expansions to other browsers. We utilized \textit{Playwright} for installing extensions while emulating browser behavior. Due to the high volume of requests generated during simulation, we employed a webpage record and replay tool named \textit{Catapult}~\cite{catapult} to minimize redundant requests.

For testing all malicious extensions in \vvv, a conversion process from V2 is necessary. This conversion utilizes a blend of pre-existing tools and our own modifications to address specific cases not covered by standard tools. The foundation of our conversion process is the \textit{extension\ manifest\ converter} tool~\cite{autoconverter}, which translates both the manifest and JS files. Key manifest fields we convert include \textit{host\_permissions}, \textit{content\_security\_policy}, \textit{background.script}, and \textit{sandbox}. Additionally, for APIs, we adapt \textit{browser\_action} to \textit{action}, \textit{tabs.executeScript} to \textit{scripting}, and modify \textit{tabs.insertCSS}.

\begin{figure}
    \centering
    \centering
    \includegraphics[width=\linewidth]{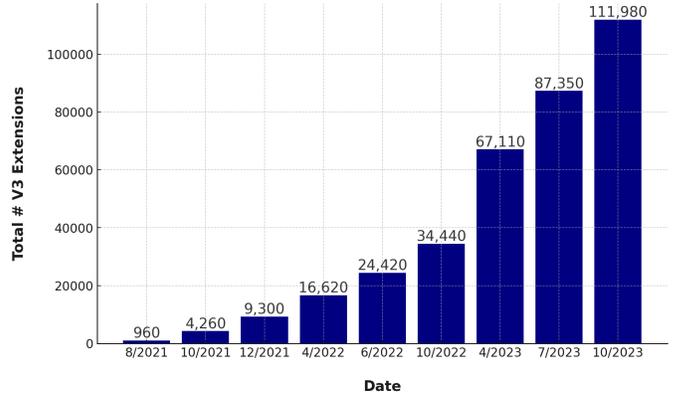}
    \caption{Number of V3 extensions in our database over time}
    \label{fig:v2v3timeseries}
\end{figure}

\section{Malicious Extension Dataset} \label{sec:data}

\subsection{Extension Sources from Past Reports} \label{sec:data:gathering1}

To create our dataset, we source extensions from various origins, predominantly security companies and previous scientific studies. Our collection focuses on malicious extensions, categorized based on their reported activities. The main source of our dataset is a resource gathered from previous malicious extension reports from the past seven years~\cite{malorybowes}. This collection comprises hashes of extensions organized based on significant incidents within the malicious ecosystem. This collection of reports offers information on malicious extension packages and has been utilized in prior research~\cite{usesmalorybowes}.

These reports include extensions stealing user data~\cite{zdnet_userdata, bleeping_analytics}, acquiring Facebook information~\cite{zdnet_facebook_sue}, capturing sensitive data~\cite{gigamon, radware}, spying on users~\cite{dataspii}, compromising credentials~\cite{zdnet_cryptowallet, medium_cryptowallet}, exploiting Chrome features~\cite{bleeping_abuse_chrome}, and bypassing authentication~\cite{ghacks_authenticator, reddit_authenticator}. Noteworthy in our collection is \textit{The Great Suspender}, implicated in user data theft~\cite{great_suspender_1, great_suspender_2}, and extensions targeting accounts and passwords~\cite{arstechnica_accounts, kaspersky_malware}, including those from foreign entities like North Korea~\cite{northkorean_malware}.




\subsection{Our Historical Dataset} \label{sec:data:gathering2}

Although the resource provides access to reports on malicious extensions, it lacks the source code. Therefore, we maintain a historical dataset of extension source codes. To compile this dataset, we download all extensions from the webstore that were updated in the last 24 hours, every day. Over six years, this method has enabled us to gather a comprehensive dataset that includes a total of \extensionsOurDataset\ extensions. This dataset serves as a valuable tool for conducting detailed analyses, queries, and comparisons between various versions of extensions.

Using this historical dataset, we have been able to identify malicious versions for \maliciousExtensions\ IDs. Our dataset is the main source for identifying 85\% of these malicious extensions. The remaining 15\% are identified with the help of the \textit{chrome-stats.com} tool~\cite{chromeStats}, which assists in finding versions that our dataset does not cover.

\section{Methodology} \label{sec:methodology}

\subsection{Architecture} \label{sec:methodology:architecture}

In Figure~\ref{fig:newArchitecture}, we present the complete architecture and methodology used in this study. The process starts with the collection of malicious data in the Manifest V2 format. We then develop a converter to transition these extensions to the Manifest V3 compatible ecosystem. After the conversion, we run each extension through a dynamic analysis tool. This step ensures the conversion's success and checks if the default settings can monitor the malicious behavior. We run it on URLs the extension has access to run on to ensure functionality. Finally, if malicious behavior exists in the Manifest V3 ecosystem, we perform a manual verification.

\subsection{Gathering \& Verifying the Malicious Datasets} \label{sec:methodology:prevAttacks}

As outlined in the data analysis section (\S~\ref{sec:data:gathering1}), due to the scarcity of reports on malicious Manifest V3 extensions, our investigation centers on those developed in Manifest V2. Initially, we merge our two partial datasets to create the final dataset. To analyze these extensions, we utilize \textit{Playwright} for automating a browser environment. This approach facilitates automated tests, with each test involving a predefined browser session featuring one of the extensions. During each test, the extensions are navigated through a series of predetermined \textit{URLs} they are allowed to run on based on information from the \textit{manifest.json} configuration file.

Following this process, we undertake the verification of the malicious behavior in all 517 extensions identified as suspicious. This verification process is twofold: we manually examine the code where possible, and in instances where the report lacks details about the malicious behavior, we dynamically execute and monitor the extension for any signs of potentially unwanted behavior. This comprehensive approach ensures a thorough assessment of each extension's activities and functions.

\begin{figure}[t]
	\centering
	\includegraphics[width=\linewidth]{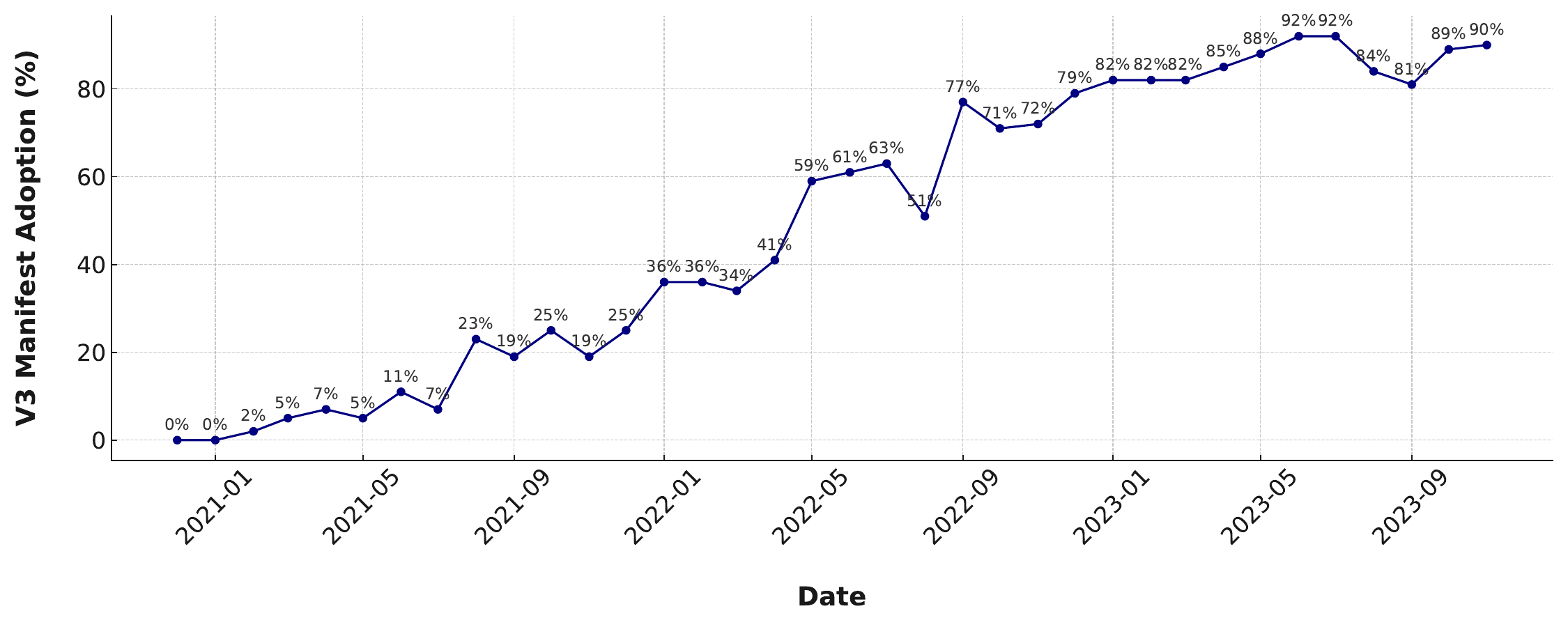}
	\caption{Manifest V3 monthly update rate over the past four years to showcase how fast the adoption is happening.}
	\label{fig:adoptionRate4Years}
\end{figure}

\subsection{Malicious Categories Labeling} \label{sec:methodology:labelling}

We categorize malicious extensions into six types to understand their behaviors and persistence in Manifest V3. These include click scams, ad replacement, user data analytics, credentials stealing, browser modification, and \textit{other} activities like crypto theft and phishing. User data analytics is the most common, capturing 68\% of labels, indicating extensive personal information collection. Browser modifications and ad replacements are also significant, each at 21\%, while credentials stealing is less common with eight instances, reflecting its high impact and execution complexity. The distribution across categories is detailed in Table~\ref{tab:malCategories1}, noting some extensions fall into multiple categories.

\subsubsection*{Manual Verification}

The manual verification process involves spending 15 minutes per extension to examine the code, culminating in approximately 130 hours of manual review.

\begin{table}[t]
	\centering
	\begin{tabular}{c|c} \hline
		\textbf{(OLD) Manifest V2}                & \textbf{(NEW) Manifest V3}         \\ \hline
		manifest\_version: 2                      & manifest\_version: 3               \\ \hline
		background\_scripts                       & service\_workers                   \\ \hline
		chrome.browseraction                      & chrome.action                      \\ \hline
		CSP policies                              & CSP policies                       \\
		(any)                                     & (self, none, localhost, 127.0.0.1) \\ \hline
		chrome.extension.*~\cite{deprecated_apis} & chrome.runtime.*                   \\ \hline
		chrome.tabs.*~\cite{deprecated_apis}      & chrome.[VARIES].*                  \\ \hline
	\end{tabular}
	\caption{Categories of Manifest V3 changes our autoconverter handles successfully.}
	\label{tab:autoconverted}
\end{table}

\begin{table}[b]
	\centering
	\begin{minipage}{.5\textwidth}
		\centering
		{\small
			\begin{tabular}{c|c} \hline
				\textbf{webstore ID} & \textbf{\# Extensions (\%)} \\ \hline
				Online               & 2,649 (84.7\%)              \\ \hline
				Offline              & 479 (15.3\%)                \\ \hline
				\textbf{Total}       & \textbf{3,128 (100\%)}      \\ \hline
			\end{tabular}
		}
		\caption{Extensions that were rolled-back from V3 back to V2 and online availability in the webstore.}
		\label{tab:rollback}
	\end{minipage}
\end{table}

\subsection{V3 Conversion} \label{sec:methodology:conversion}

In the process of adapting extensions from Manifest Version 2 to Version 3, our autoconverter, which is based on official documentation, focuses on crucial conversions as outlined in Table~\ref{tab:autoconverted}. The conversion process updates the \textit{manifest.json} file, which is essential for the functionality of the extension, as well as deprecated APIs in the source code of the extension. The \textit{manifest.json} changes include updating the \textit{manifest\_version} to 3 to indicate compliance with the most recent specifications. Background scripts are replaced with \textit{service\_workers} to align with Manifest Version 3's emphasis on more efficient and secure background processes. Furthermore, the transition from \textit{chrome.browseraction} to \textit{chrome.action} aligns with the new Manifest standards, and Content Security Policy (CSP) policies are updated to values accepted by Manifest Version 3.

The converter also addresses the deprecation of certain APIs by transitioning from the deprecated \textit{chrome.extension.*} to \textit{chrome.runtime.*}, for the subset of APIs affected and reported in the official website~\cite{deprecated_apis}. Nevertheless, our converter encounters challenges with extensions that rely heavily on background scripts for interacting with the Document Object Model (DOM) or require complex modifications to the \textit{manifest.json} file, such as adapting scripts to work with service workers, which results to our converter not providing 100\% coverage for all testcases. Furthermore, we note an open-source project known as the extension manifest converter~\cite{autoconverter} which takes some of these conversions into account. However, this project is not actively maintained and supports a more limited range of APIs compared to our autoconverter, highlighting the more comprehensive scope and enhanced capabilities of our tool.

\subsubsection*{Benign vs Malicious Extensions}

The conversion process, including manifest and API changes, applies equally to both benign and malicious extensions, ensuring that both types have the same likelihood of functioning after conversion. This is due to a broad set of changes affecting all extensions regardless of their intent. The neutrality of our conversion process means that the outcome for each extension, whether benign or malicious, is determined by its compatibility with Manifest Version 3 specifications, not its original purpose.

\begin{figure}[tb]
	\centering
	\includegraphics[width=\linewidth]{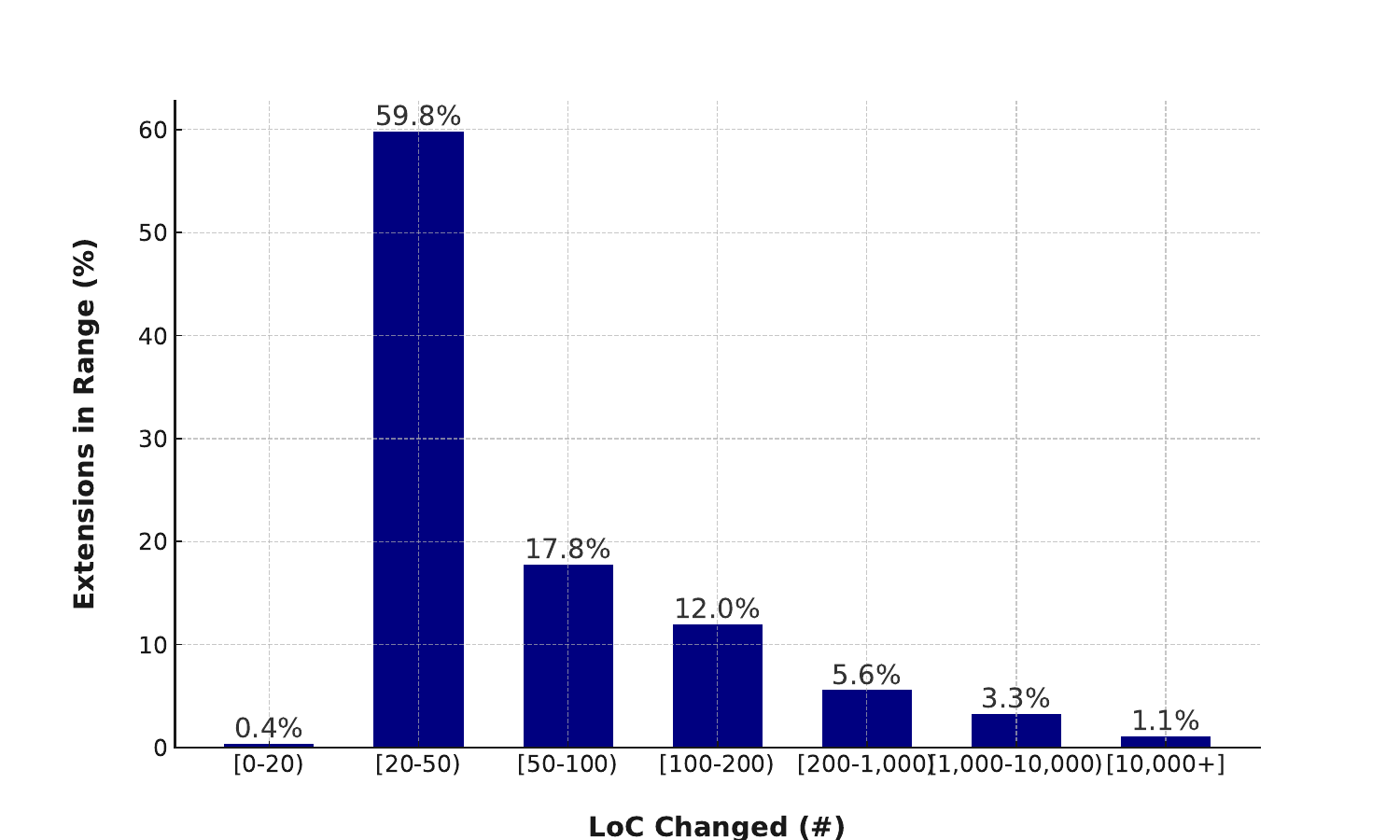}
	\caption{How many LoC needed to convert the V2 malicious extensions to a V3 equivalent version before running the dynamic analysis on the V3 ones.}
	\label{fig:locDistribution}
\end{figure}

\subsection{Dynamic Analysis}

To test if a malicious extension remains functional after conversion, we employ playwright instrumentation to load the extension in a Chromium browser. We then navigate to websites allowed by the extension's manifest.json configuration. Successful website visits indicate that our conversion process has preserved the extension's functionality.

\subsection{Malicious Extensions Definition} \label{sec:methodology:maliciousness}

In our final methodology phase, we dynamically test converted V2 extensions in the Manifest V3 framework to assess their active status. We use \textit{Catapult} for secure session recording and replay, preventing direct external server communication. By comparing sessions with and without the extension, we identify extension-specific requests and check them against a list of known malicious \textit{URLs}, including \textit{C\&C} servers and domains listed on \textit{EasyList} and \textit{EasyPrivacy}~\cite{easylist}.




\subsubsection*{Functionally Active} We categorize an extension as a functionally active malicious extension in Manifest V3 when it meets specific criteria: it has a history of verified malicious behavior, has been removed from the Google Webstore, falls into one of the malicious categories after manual verification of its behavior, successfully converts to Manifest V3 and loads correctly, and attempts to initiate a request to any \textit{URL} from a list of known malicious domains. These criteria are vital for assessing whether extensions, previously recognized as malicious in their V2 form, continue to operate maliciously under Manifest V3.

\section{Results} \label{sec:results}

\subsection{Manifest V3 Adoption Rate} \label{sec:results:evolution}

The analysis of Manifest V3 adoption in the Chrome Webstore highlights a steady transition from V2 to V3. Initially, the V3 conversion rate was below 5\%, with only about 30,000 extension versions updated to V3. This trend saw a significant increase, with 90\% of new uploads being in V3 by late 2023/early 2024, according to the data depicted in Figure~\ref{fig:v2v3timeseries} and Figure~\ref{fig:adoptionRate4Years}. These figures illustrate the monthly adoption rates, showing a consistent rise in V3 adoption from 2\% in January 2021 to 40\% of new daily updates by the time the migration plan was paused.

Furthermore, our findings on rollback rates, where extensions reverted to V2 after initially updating to V3, involve 3,129 extensions. Table~\ref{tab:rollback} details that 15.3\% of these rolled-back extensions were removed from the Webstore, indicating challenges with V3 adoption due to its limitations or concerns over malicious behavior. This comprehensive analysis underscores a significant yet gradual shift towards Manifest V3, marked by initial hesitancy and eventual widespread acceptance among developers.

\begin{table}[t]
  \begin{minipage}{.5\textwidth}
    \centering
    {\small
      \begin{tabular}{c|c} \hline
        \textbf{Result}                   & \textbf{Number of extensions (\%)} \\ \hline
        Success (Initial)                 & 154 (29.8\%)                       \\ \hline
        Fail (Initial)                    & 363 (70.2\%)                       \\ \hline
        \hline
        Success (after war modifications) & 290 (56.1\%)                       \\ \hline
        Fail (Final)                      & 227 (43.9\%)                       \\ \hline
        \hline
        \textbf{Executed}                 & \textbf{517 (100\%)}               \\ \hline
      \end{tabular}
    }
    \caption{Outcome of dynamic execution of malicious V3 extensions}
    \label{tab:dynamic1}
  \end{minipage}
\end{table}

\begin{table*}[t]
  \centering
  {\small
    \begin{tabular}{c|c|c|c}
      \hline
      \textbf{API}          & \textbf{API}                  & \textbf{Vulnerability} & \textbf{Malicious code} \\
      \textbf{Category}     & \textbf{Name}                 & \textbf{Related}       & \textbf{Related}        \\
      \hline
                            & runtime.sendMessage           & \ding{51}              & \ding{55}               \\
      Background Pages      & runtime.connect               & \ding{51}              & \ding{55}               \\
      Related APIs          & runtime.onMessage.addListener & \ding{51}              & \ding{55}               \\
                            & runtime.onConnect.addListener & \ding{51}              & \ding{55}               \\
      \hline
      Web Request           & webRequest                    & \ding{51}              & \ding{51}               \\
      API                   & webRequestBlocking            & \ding{51}              & \ding{51}               \\
      \hline
      Content Scripts and   & XMLHttpRequest                & \ding{51}              & \ding{51}               \\
      Cross-Origin Requests & fetch                         & \ding{51}              & \ding{51}               \\
      \hline
      Remotely Hosted Code  & eval                          & \ding{55}              & \ding{51}               \\
      \hline
    \end{tabular}
  }
  \caption{APIs that were deprecated, replaced or offered alternatives in Manifest V3 which are also related to vulnerable and malicious JavaScript code in browser extensions.}
  \label{tab:malvulnapis}
\end{table*}

\begin{table*}[t]
  \centering
  {\small
    \begin{tabular}{c|c|c|c|c}
      \hline
      \textbf{API}              & \textbf{API}                  & \textbf{Total API} & \textbf{Unique Extension} & \textbf{API exists} \\
      \textbf{Category}         & \textbf{Name}                 & \textbf{Hits}      & \textbf{Hits}             & \textbf{(\%)}       \\
      \hline
                                & runtime.sendMessage           & 1,386              & 112                       & 21.7                \\
      Background Pages          & runtime.connect               & 242                & 69                        & 13.3                \\
      Related APIs              & runtime.onMessage.addListener & 874                & 88                        & 17.0                \\
                                & runtime.onConnect.addListener & 50                 & 21                        & 4.1                 \\
      \hline
      Web Request               & webRequest                    & 1,004              & 85                        & 16.4                \\
      API                       & webRequestBlocking            & 234                & 35                        & 6.8                 \\
      \hline
      Content Scripts and       & XMLHttpRequest                & 4,340              & 340                       & 65.8                \\
      Cross-Origin Requests     & fetch                         & 3,972              & 312                       & 60.3                \\
      \hline
      Remotely Hosted Code      & eval                          & 6,654              & 454                       & 87.8                \\
      \hline
      \textbf{Total Extensions} & \textbf{N/A}                  & \textbf{N/A}       & \textbf{517}              & \textbf{100}        \\
      \hline
    \end{tabular}
  }
  \caption{Total hits of APIs related to vulnerable and malicious code in dynamically tested malicious extensions.}
  \label{tab:malvulnhits}
\end{table*}

\subsection{LoC Distribution after Conversion to V3} \label{sec:results:loc}

The conversion of malicious V2 extensions to Manifest V3 showed varied changes in lines of code (LoC), which we detail in Figure~\ref{fig:locDistribution}. Notably, conversions rarely involved fewer than 20 changed LoC. The majority, 89.6\%, saw changes ranging from 20 to 200 LoC. Outliers, constituting 1.1\% of the dataset, experienced changes exceeding 10,000 LoC, often in extensions with large source code bundles. This highlights the diverse effects of transitioning to V3, influenced by the extension's complexity and code structure.



The significant changes in lines of code are attributed to the size of libraries included in the extension source code and our modification process. Initially, we substitute APIs in the code, followed by code beautification to facilitate line count. If a large library contains approximately 10,000 lines of code on a single line and we perform a substitution there, then beautify it, this results in it being counted as 10,000 lines of code changed.

\subsection{Improved Security of Manifest V3}

Manifest V3 has improved security by altering API usage, with changes detailed in Table~\ref{tab:malvulnapis}, informed by prior work on vulnerable and malicious extensions~\cite{browserX, pantelaios_malicious}. These changes target APIs previously linked to vulnerabilities or malicious activities. Table~\ref{tab:malvulnhits} shows the usage of such APIs in malicious extensions, highlighting the impact of these modifications. For example, the \textit{eval} function, highly susceptible to misuse, and APIs like \textit{XMLHttpRequest} and \textit{fetch}, are extensively used in malicious activities, as evidenced by their high usage rates.

\subsection{Dynamic Execution for V3 Malicious Extensions} \label{sec:results:dynamic}

In our analysis, we dynamically tested all \maliciousExtensions\ extensions converted to Manifest V3, with \stillRunFirst\ exhibiting malicious behavior post-conversion, representing \stillRunFirstPercentage\ of the tested set. Results in Table~\ref{tab:dynamic1} reveal \notStillRunFirstPercentage\ of extensions failed to display malicious activity due to \war\ or DOM issues. Table~\ref{tab:dynamic2} categorizes the behaviors of extensions that ran successfully, showing user data analytics and browser modifications as dominant categories, as initially categorized in Table~\ref{tab:malCategories1}. The comprehensive results of the extension conversion process, post implementation of the proof of concept, are detailed in Table~\ref{tab:autoconverted}. This table reveals that 56.1\% of the total extensions retained functionality after conversion to V3 and the application of the proof of concept.

\subsection{Proof of Concept for Code Injection} \label{sec:results:war}

Manifest V3 changes how third-party code is included in extensions, requiring executable code to be part of the source code, yet allowing for external resource interaction through \war\ files declared in the \textit{manifest.json}. This setup enables extensions to integrate third-party scripts, potentially creating a chain of JavaScript file inclusions. Listings~\ref{code:3rdparty_1} and~\ref{code:3rdparty_2} provide a proof of concept, showing the declaration for external interactions and how remote server files can inject further JS scripts. This method opens avenues for various malicious activities, including session cookie theft and password compromises.



\section{Discussion \& Mitigations} \label{sec:discussion}

\begin{table}[b]
  \centering
  {\small
    \begin{tabular}{c|c} \hline
      \textbf{Malicious label} & \textbf{Number of functional extensions} \\ \hline
      Click scam               & 15                                       \\ \hline
      Ad replacement           & 77                                       \\ \hline
      User data analytics      & 123                                      \\ \hline
      Credentials stealing     & 2                                        \\ \hline
      Browser modification     & 88                                       \\ \hline
      Other                    & 29                                       \\ \hline
    \end{tabular}
  }
  \caption{Malicious v3 extensions that ran successfully per malicious category}
  \label{tab:dynamic2}
\end{table}

\subsection{Pioneering Study \& Community Response}

This study is the first to analyze V3 adoption and its impact on malicious extensions, revealing that V3 does not eliminate any malicious category entirely. It also provides proof of concept for converting V2 extensions to V3, preserving their functionalities.

Manifest V3's rollout has sparked debate, especially among adblocker developers concerned about reduced efficacy~\cite{ghostery_1, negative_publicity_1}. The transition to the \textit{declarativeNetRequest (DNR)} API from \textit{webRequest} has raised issues about extension capability and innovation~\cite{negative_publicity_2, negative_publicity_3, negative_publicity_4, negative_publicity_5}, despite some studies indicating potential performance benefits for privacy extensions~\cite{ghostery_study, privacy_study}. Challenges with service workers' DOM API access and library compatibility further highlight the need for V3 improvements.

\subsection{Positives of V3 \& Mitigation Strategies}

Manifest V3 has introduced measures enhancing extension security, notably making malicious XMLHttpRequest (XHR) and fetch requests harder to execute by restricting XHR to content scripts and requiring Cross-Origin Resource Sharing (CORS) headers adjustment~\cite{cors_updated}. The review process for extensions is also being improved~\cite{mitigation_claims_1, mitigation_claims_2, mitigation_claims_3, mitigation_claims_4}, raising security standards.

To further mitigate risks, strategies include deploying 3rd-party watchers to monitor \war\ files for malicious changes, dynamically executing extensions to identify unauthorized redirects. Adjusting Content Security Policy (CSP) rules to limit malicious extensions without affecting benign ones and enhancing manual reviews coupled with advocating for open-source code to allow community-driven security checks are proposed, building on top of tools that analyze extensions already like \textit{CRXcavator}~\cite{crxcavatorOriginal} and \textit{LayerX}~\cite{layerx}. These strategies, although challenging due to potential evasion techniques~\cite{jordan_residential}, are crucial for enhancing V3 security and ensuring a safer extension ecosystem.

\begin{lstlisting}[style=ES6, frame=tb, caption=3rd party inclusion - manifest.json, captionpos=b, belowcaptionskip=-1cm, label=code:3rdparty_1]
  // manifest.json
  [..]
  "content_security_policy": {
    "extension_pages": "script-src 'self'; object-src 'self'"},
  "web_accessible_resources": [
    {"resources": ["src/injects_3rd_party.js"],
      "matches": ["https://*/*"]}]
  [..]
  \end{lstlisting}

\begin{lstlisting}[style=ES6, frame=tb, caption=Inclusion of 3rd party: (a) local extension file and (b) external payload, captionpos=b, belowcaptionskip=-1cm, label=code:3rdparty_2]
  // (a) injects_3rd_party.js
  (function (e) {
      var site = "//malicious_site.com";
      var script = e.createElement("script");
      script.src = "https:" + site + "/js/malicious_3rd_party_payload.js";
      (e.head || e.body).appendChild(script)
  }) (window, document);
  --------------------------
  // (b) malicious_3rd_party_payload.js
  do_malicious_stuff()
  \end{lstlisting}

\section{Limitations \& Future Work} \label{sec:limitations}

\subsection{Evasion Techniques in Malicious Extensions} \label{sec:limitations:evasion}
Our evaluation of malicious extensions in V3 highlighted the impact of evasion techniques on our ability to detect malicious behaviors, as detailed in Table~\ref{tab:evasions}. These techniques range from monitoring user actions to delaying malicious payloads, showcasing the adaptability of malicious actors. Addressing these sophisticated evasion tactics is crucial for enhancing security measures.

\subsection{Future work} \label{sec:limitations:futurework}
Future efforts will broaden our analysis to additional browsers like Firefox, Opera, Safari, and Edge, focusing on the adaptation to \vvv\ and service workers. Gathering a comprehensive dataset of malicious extensions across these platforms presents a significant challenge. We also aim to assess the V3 ecosystem's evolution post-V2 phase-out, especially its impact on the extension landscape and security measures.

\begin{table}
    \centering
    {\small
        \begin{tabular}{c} \hline
            \textbf{Evasion techniques}                             \\ \hline
            Deactivate if IP is from a university                   \\ \hline
            Deactivate if user's search query contains the words    \\ "C\&C" or other similar phrases \\ \hline
            Deactivate if user accesses locally hosted websites     \\ \hline
            Wait 3 days until malicious payload is downloaded       \\ \hline
            Check against a hard-coded list of privacy extensions   \\ \hline
            Check whether user is accessing the Developer Tools API \\ \hline
            Check if user is tech-savvy (combination of above)      \\ \hline
        \end{tabular}
    }
    \caption{Evasion techniques used by malicious extensions}
    \label{tab:evasions}
\end{table}




\section{Related Work} \label{sec:relatedwork}

The extension ecosystem, active for over a decade, has been extensively researched by the scientific community.

A significant portion of this research focuses on identifying and analyzing malicious extensions. Kapravelos et al.~\cite{hulk_kapravelos} developed a method to detect such extensions using honey pages. Weissbacher et al.~\cite{weissbacher_history} investigated extensions leaking browser history, whereas Chen et al.~\cite{mystique} performed analysis based on data sources and sinks in extensions. Starov et al.~\cite{starov_diffusion} examined privacy diffusion in extensions, creating a tool named \textit{PrivacyMeter} for this purpose. Research by Pantelaios et al.~\cite{pantelaios_malicious} involved identifying malicious extensions through clustering of similar JS API changes.

In the realm of privacy preservation and vulnerable data flows, Fass et al.~\cite{browserX} introduced \textit{DoubleX}, a tool for detecting vulnerable data flows generated by extensions. Zhao et al.~\cite{zhao_chinese} focused specifically on privacy leaks in Chinese extensions. Starov et al.~\cite{starov_privacy} created an extension dedicated to privacy preservation. The \textit{Empoweb} tool, aimed at identifying APIs used by vulnerable extensions, is another notable contribution in this area~\cite{empoweb}. Giuffrida et al.~\cite{giuffrida_crossbrowser} developed a model for detecting privacy breaches in cross-browser extensions, and Li et al.~\cite{li_spyshield} introduced \textit{SpyShield}, a tool for preserving privacy in add-ons.

The study by Borgolte et al.~\cite{privacy_study} highlighted the performance benefits of privacy-focused extensions. This work was cited by ad blocker developers criticizing the limitations of \vvv~\cite{ghostery_1}. Xie et al.~\cite{xie_jtaint} investigated privacy leaks in Chrome extensions using \textit{JTaint}, a JavaScript analysis tool. Zhu et al.~\cite{zhu_shadowblock} developed a lightweight, stealthy adblocking browser. Finally, Agarwal et al. identified that more than 2,400 extensions interfere with security headers in the domain of extension security headers~\cite{Agarwal2021FirstDN}.

\section{Conclusion} \label{sec:conclusion}

Manifest V3 significantly enhances Chrome extension security by deprecating or removing 87.8\% of vulnerable APIs. Despite this, \stillRunFirst\ (\stillRunFirstPercentage) of analyzed malicious extensions remain operational after conversion. Initially, less than 5\% of extensions adopted V3, but now 90\% of new uploads are in V3. Our analysis of 517 malicious extensions indicates a reduction in functionality post-V3, yet \stillRunSecond\ (\stillRunSecondPercentage) adapt using web accessible resources for malicious activities. These findings underline the ongoing need for improvements in the extension ecosystem.

\bibliographystyle{IEEEtranS}

\bibliography{bibliography}

\begin{thebibliography}{10}
\providecommand{\url}[1]{#1}
\csname url@samestyle\endcsname
\providecommand{\newblock}{\relax}
\providecommand{\bibinfo}[2]{#2}
\providecommand{\BIBentrySTDinterwordspacing}{\spaceskip=0pt\relax}
\providecommand{\BIBentryALTinterwordstretchfactor}{4}
\providecommand{\BIBentryALTinterwordspacing}{\spaceskip=\fontdimen2\font plus
\BIBentryALTinterwordstretchfactor\fontdimen3\font minus
  \fontdimen4\font\relax}
\providecommand{\BIBforeignlanguage}[2]{{%
\expandafter\ifx\csname l@#1\endcsname\relax
\typeout{** WARNING: IEEEtranS.bst: No hyphenation pattern has been}%
\typeout{** loaded for the language `#1'. Using the pattern for}%
\typeout{** the default language instead.}%
\else
\language=\csname l@#1\endcsname
\fi
#2}}
\providecommand{\BIBdecl}{\relax}
\BIBdecl

\bibitem{manifestv3_1}
``Manifest v3,'' \url{https://developer.chrome.com/docs/extensions/mv3/intro/},
  2020.

\bibitem{declarativeNetRequest}
``declarativenetrequest api,''
  \url{https://developer.chrome.com/docs/extensions/reference/declarativeNetRequest/},
  2021.

\bibitem{service_workers}
``Service worker api,''
  \url{https://developer.chrome.com/docs/workbox/service-worker-overview/},
  2021.

\bibitem{ghostery_study}
``Adblocker performance study by ghostery,''
  \url{https://whotracks.me/blog/adblockers_performance_study.html}, 2022.

\bibitem{apiDocumentation}
``Api deprecation in v3,''
  \url{https://developer.chrome.com/docs/extensions/mv3/intro/mv3-migration/},
  2022.

\bibitem{arstechnica_accounts}
``arstechnica: extesions stealing user accounts,''
  \url{https://arstechnica.com/information-technology/2020/10/popular-chromium-ad-blockers-caught-stealing-user-data-and-accessing-accounts/},
  2022.

\bibitem{bleeping_abuse_chrome}
``bleepingcomputer: extension abusing chrome,''
  \url{https://www.bleepingcomputer.com/news/security/malicious-extension-abuses-chrome-sync-to-steal-users-data/},
  2022.

\bibitem{bleeping_analytics}
``bleepingcomputer: extesions stealing user analytics,''
  \url{https://www.bleepingcomputer.com/news/security/facebook-sues-makers-of-malicious-chrome-extensions-for-scraping-data/},
  2022.

\bibitem{blockSite}
``Blocksite extension,''
  \url{https://chrome.google.com/webstore/detail/blocksite-block-websites/eiimnmioipafcokbfikbljfdeojpcgbh},
  2022.

\bibitem{mitigation_claims_3}
``Browser v3 adoption,''
  \url{https://extensionworkshop.com/documentation/develop/manifest-v3-migration-guide/},
  2022.

\bibitem{mitigation_claims_1}
``Browsers collaborating with developers,''
  \url{https://blog.mozilla.org/addons/2022/05/18/manifest-v3-in-firefox-recap-next-steps/},
  2022.

\bibitem{negative_publicity_2}
``Browser's existing policies solutions,''
  \url{https://www.eff.org/deeplinks/2019/07/googles-plans-chrome-extensions-wont-really-help-security},
  2022.

\bibitem{catapult}
``Catapult web page replay,''
  \url{https://chromium.googlesource.com/catapult/+/HEAD/web_page_replay_go/README.md},
  2022.

\bibitem{chromeStats}
``chrome-stats.com,'' \url{https://chrome-stats.com/}, 2022.

\bibitem{cors_updated}
``Cors header updated in chrome v85,''
  \url{https://www.chromium.org/Home/chromium-security/extension-content-script-fetches/},
  2022.

\bibitem{dataspii}
``dataspii: extensions stealing user information,''
  \url{https://www.salon.com/2019/07/22/malicious-browser-extensions-are-stealing-personal-information/},
  2022.

\bibitem{content_script}
``Extension content scripts,''
  \url{https://developer.chrome.com/docs/extensions/mv3/content_scripts/},
  2022.

\bibitem{northkorean_malware}
``Extension malware from north korea,''
  \url{https://www.cisa.gov/uscert/ncas/alerts/aa20-301a}, 2022.

\bibitem{autoconverter}
``Extension manifest converter,''
  \url{https://github.com/GoogleChromeLabs/extension-manifest-converter}, 2022.

\bibitem{negative_publicity_1}
``Extensions' potential malfunction article,''
  \url{https://www.techrepublic.com/article/google-makes-the-perfect-case-for-why-you-shouldnt-use-chrome/},
  2022.

\bibitem{mitigation_claims_2}
``Firefox collaboration on extensions,''
  \url{https://blog.mozilla.org/addons/2021/05/27/manifest-v3-update/}, 2022.

\bibitem{ghacks_authenticator}
``ghacks: extension faking authenticator,''
  \url{https://www.ghacks.net/2021/05/18/dont-download-this-microsoft-authenticator-extension-for-chrome-it-is-fake/},
  2022.

\bibitem{ghostery_1}
``Ghostery extension report,''
  \url{https://www.ghostery.com/blog/manifest-v3-the-ghostery-perspective},
  2022.

\bibitem{gigamon}
``Gigamon: extensions with criminal implications,''
  \url{https://blog.gigamon.com/2018/01/18/malicious-chrome-extensions-enable-criminals-to-impact-half-a-million-users-and-global-businesses/},
  2022.

\bibitem{kaspersky_malware}
``kaspersky: extensions stealing user data,''
  \url{https://www.kaspersky.com/blog/chrome-plugins-alert/38242/}, 2022.

\bibitem{passwordManager}
``Lastpass extension,''
  \url{https://chrome.google.com/webstore/detail/lastpass-free-password-ma/hdokiejnpimakedhajhdlcegeplioahd},
  2022.

\bibitem{great_suspender_1}
``lifehacker: great suspender extension,''
  \url{https://lifehacker.com/ditch-the-great-suspender-before-it-becomes-a-security-1845989664},
  2022.

\bibitem{negative_publicity_3}
``Manifest v3 can hurt innocation,''
  \url{https://www.eff.org/deeplinks/2021/12/googles-manifest-v3-still-hurts-privacy-security-innovation},
  2022.

\bibitem{medium_cryptowallet}
``medium: extensions stealing cryptowallet data,''
  \url{https://medium.com/mycrypto/discovering-fake-browser-extensions-that-target-users-of-ledger-trezor-mew-metamask-and-more-e281a2b80ff9},
  2022.

\bibitem{mitigation_claims_4}
``Microsoft v3 rollout,''
  \url{https://docs.microsoft.com/en-us/microsoft-edge/extensions-chromium/developer-guide/manifest-v3},
  2022.

\bibitem{playwright}
``Playwright browser simulation,'' \url{https://playwright.dev/}, 2022.

\bibitem{negative_publicity_4}
``Potential threats of v3,''
  \url{https://www.eff.org/deeplinks/2021/12/chrome-users-beware-manifest-v3-deceitful-and-threatening},
  2022.

\bibitem{negative_publicity_5}
``Privacy extensions changes,''
  \url{https://www.theregister.com/2022/06/08/google_blocking_privacy_manifest/},
  2022.

\bibitem{radware}
``radware: crypto stealing extensions,''
  \url{https://blog.radware.com/security/2018/05/nigelthorn-malware-abuses-chrome-extensions/},
  2022.

\bibitem{reddit_authenticator}
``reddit: extension faking authenticator,''
  \url{https://www.reddit.com/r/chrome/comments/hbpi7z/found_a_extension_that_contains_malware/},
  2022.

\bibitem{unblockOrigin}
``ublock origin extension,''
  \url{https://chrome.google.com/webstore/detail/ublock-origin/cjpalhdlnbpafiamejdnhcphjbkeiagm},
  2022.

\bibitem{great_suspender_2}
``xda-devlopers: great suspender extension,''
  \url{https://www.xda-developers.com/google-chrome-the-great-suspender-malware/},
  2022.

\bibitem{zdnet_cryptowallet}
``zdnet: extensions stealing cryptowallet private keys,''
  \url{https://www.zdnet.com/article/chrome-extension-caught-stealing-crypto-wallet-private-keys/},
  2022.

\bibitem{zdnet_userdata}
``zdnet: extensions stealing sensitive user data,''
  \url{https://www.zdnet.com/article/google-removes-106-chrome-extensions-for-collecting-sensitive-user-data/},
  2022.

\bibitem{zdnet_facebook_sue}
``zdnet: facebook suing extension developers,''
  \url{https://www.zdnet.com/article/facebook-sues-two-chrome-extension-makers-for-scraping-user-data/},
  2022.

\bibitem{malorybowes}
``Collection of malicious extensions report,''
  \url{https://github.com/mallorybowes/chrome-mal-ids}, 2024.

\bibitem{crxcavatorOriginal}
``Crxcavator: Code analyzer,'' \url{https://crxcavator.io/}, 2024.

\bibitem{easylist}
``Easylist blocklisted urls,'' \url{https://github.com/easylist/easylist},
  2024.

\bibitem{layerx}
``Layerx: Code analyzer,'' \url{https://layerxsecurity.com/}, 2024.

\bibitem{deprecated_apis}
``Unsupported apis,''
  \url{https://developer.chrome.com/docs/extensions/develop/migrate/api-calls#replace-unsupported-apis},
  2024.

\bibitem{usesmalorybowes}
\BIBentryALTinterwordspacing
S.~Agarwal, ``Helping or hindering? how browser extensions undermine
  security,'' in \emph{Proceedings of the 2022 ACM SIGSAC Conference on
  Computer and Communications Security}, ser. CCS '22.\hskip 1em plus 0.5em
  minus 0.4em\relax New York, NY, USA: Association for Computing Machinery,
  2022, p. 23–37. [Online]. Available:
  \url{https://doi.org/10.1145/3548606.3560685}
\BIBentrySTDinterwordspacing

\bibitem{Agarwal2021FirstDN}
\BIBentryALTinterwordspacing
S.~Agarwal and B.~Stock, ``First, do no harm: Studying the manipulation of
  security headers in browser extensions,'' \emph{Proceedings 2021 Workshop on
  Measurements, Attacks, and Defenses for the Web}, 2021. [Online]. Available:
  \url{https://api.semanticscholar.org/CorpusID:233288141}
\BIBentrySTDinterwordspacing

\bibitem{privacy_study}
\BIBentryALTinterwordspacing
K.~Borgolte and N.~Feamster, ``{Understanding The Performance Costs and
  Benefits of Privacy-focused Browser Extensions},'' in \emph{Proceedings of
  the 29th The Web Conference (TheWebConf, formerly known as WWW)}, T.-Y. Liu
  and M.~van Steen, Eds.\hskip 1em plus 0.5em minus 0.4em\relax International
  World Wide Web Conference Committee (IW3C2), 2020. [Online]. Available:
  \url{http://dx.doi.org/10.1145/3366423.3380292}
\BIBentrySTDinterwordspacing

\bibitem{mystique}
\BIBentryALTinterwordspacing
Q.~Chen and A.~Kapravelos, ``Mystique: Uncovering information leakage from
  browser extensions,'' in \emph{Proceedings of the 2018 ACM SIGSAC Conference
  on Computer and Communications Security}, ser. CCS '18.\hskip 1em plus 0.5em
  minus 0.4em\relax New York, NY, USA: Association for Computing Machinery,
  2018, p. 1687–1700. [Online]. Available:
  \url{https://doi.org/10.1145/3243734.3243823}
\BIBentrySTDinterwordspacing

\bibitem{browserX}
\BIBentryALTinterwordspacing
A.~Fass, D.~F. Som\'{e}, M.~Backes, and B.~Stock, ``Doublex: Statically
  detecting vulnerable data flows in browser extensions at scale,'' in
  \emph{Proceedings of the 2021 ACM SIGSAC Conference on Computer and
  Communications Security}, ser. CCS '21.\hskip 1em plus 0.5em minus
  0.4em\relax New York, NY, USA: Association for Computing Machinery, 2021, p.
  1789–1804. [Online]. Available:
  \url{https://doi.org/10.1145/3460120.3484745}
\BIBentrySTDinterwordspacing

\bibitem{giuffrida_crossbrowser}
\BIBentryALTinterwordspacing
C.~Giuffrida, S.~Ortolani, and B.~Crispo, ``Memoirs of a browser: A
  cross-browser detection model for privacy-breaching extensions,'' in
  \emph{Proceedings of the 7th ACM Symposium on Information, Computer and
  Communications Security}, ser. ASIACCS '12.\hskip 1em plus 0.5em minus
  0.4em\relax New York, NY, USA: Association for Computing Machinery, 2012, p.
  10–11. [Online]. Available: \url{https://doi.org/10.1145/2414456.2414461}
\BIBentrySTDinterwordspacing

\bibitem{jordan_residential}
J.~Jueckstock, P.~Snyder, S.~Sarker, A.~Kapravelos, and B.~Livshits,
  ``{Measuring the Privacy vs. Compatibility Trade-off in Preventing
  Third-Party Stateful Tracking},'' in \emph{Proceedings of The Web Conference
  (WWW)}, Apr. 2022.

\bibitem{hulk_kapravelos}
A.~Kapravelos, C.~Grier, N.~Chachra, C.~Kruegel, G.~Vigna, and V.~Paxson,
  ``Hulk: Eliciting malicious behavior in browser extensions,'' in \emph{23rd
  USENIX Security Symposium (USENIX Security 14)}, 2014.

\bibitem{li_spyshield}
Z.~Li, X.~Wang, and J.~Y. Choi, ``Spyshield: Preserving privacy from spy
  add-ons,'' in \emph{Recent Advances in Intrusion Detection}, C.~Kruegel,
  R.~Lippmann, and A.~Clark, Eds.\hskip 1em plus 0.5em minus 0.4em\relax
  Berlin, Heidelberg: Springer Berlin Heidelberg, 2007, pp. 296--316.

\bibitem{pantelaios_malicious}
\BIBentryALTinterwordspacing
N.~Pantelaios, N.~Nikiforakis, and A.~Kapravelos, \emph{You've Changed:
  Detecting Malicious Browser Extensions through Their Update Deltas}.\hskip
  1em plus 0.5em minus 0.4em\relax New York, NY, USA: Association for Computing
  Machinery, 2020, p. 477–491. [Online]. Available:
  \url{https://doi.org/10.1145/3372297.3423343}
\BIBentrySTDinterwordspacing

\bibitem{empoweb}
D.~F. Somé, ``Empoweb: Empowering web applications with browser extensions,''
  in \emph{2019 IEEE Symposium on Security and Privacy (SP)}, 2019, pp.
  227--245.

\bibitem{starov_diffusion}
\BIBentryALTinterwordspacing
O.~Starov and N.~Nikiforakis, ``Extended tracking powers: Measuring the privacy
  diffusion enabled by browser extensions,'' in \emph{Proceedings of the 26th
  International Conference on World Wide Web}, ser. WWW '17.\hskip 1em plus
  0.5em minus 0.4em\relax Republic and Canton of Geneva, CHE: International
  World Wide Web Conferences Steering Committee, 2017, p. 1481–1490.
  [Online]. Available: \url{https://doi.org/10.1145/3038912.3052596}
\BIBentrySTDinterwordspacing

\bibitem{starov_privacy}
------, ``Privacymeter: Designing and developing a privacy-preserving browser
  extension,'' in \emph{Engineering Secure Software and Systems}, M.~Payer,
  A.~Rashid, and J.~M. Such, Eds.\hskip 1em plus 0.5em minus 0.4em\relax Cham:
  Springer International Publishing, 2018, pp. 77--95.

\bibitem{weissbacher_history}
\BIBentryALTinterwordspacing
M.~Weissbacher, E.~Mariconti, G.~Suarez-Tangil, G.~Stringhini, W.~Robertson,
  and E.~Kirda, ``Ex-ray: Detection of history-leaking browser extensions,'' in
  \emph{Proceedings of the 33rd Annual Computer Security Applications
  Conference}, ser. ACSAC '17.\hskip 1em plus 0.5em minus 0.4em\relax New York,
  NY, USA: Association for Computing Machinery, 2017, p. 590–602. [Online].
  Available: \url{https://doi.org/10.1145/3134600.3134632}
\BIBentrySTDinterwordspacing

\bibitem{xie_jtaint}
M.~Xie, J.~Fu, J.~He, C.~Luo, and G.~Peng, ``Jtaint: Finding privacy-leakage in
  chrome extensions,'' in \emph{Information Security and Privacy}, J.~K. Liu
  and H.~Cui, Eds.\hskip 1em plus 0.5em minus 0.4em\relax Cham: Springer
  International Publishing, 2020, pp. 563--583.

\bibitem{zhao_chinese}
Y.~Zhao, L.~Yang, Z.~Li, L.~He, and Y.~Zhang, ``Privacy model: Detect privacy
  leakage for chinese browser extensions,'' \emph{IEEE Access}, vol.~9, pp.
  44\,502--44\,513, 2021.

\bibitem{zhu_shadowblock}
\BIBentryALTinterwordspacing
S.~Zhu, U.~Iqbal, Z.~Wang, Z.~Qian, Z.~Shafiq, and W.~Chen, ``Shadowblock: A
  lightweight and stealthy adblocking browser,'' in \emph{The World Wide Web
  Conference}, ser. WWW '19.\hskip 1em plus 0.5em minus 0.4em\relax New York,
  NY, USA: Association for Computing Machinery, 2019, p. 2483–2493. [Online].
  Available: \url{https://doi.org/10.1145/3308558.3313558}
\BIBentrySTDinterwordspacing

\end{thebibliography}

\end{document}